\documentclass[aps,prb,twocolumn,showpacs,amsmath,amssymb,reprint,superscriptaddress]{revtex4-1}

\usepackage{graphicx}
\usepackage{dcolumn}
\usepackage{bm}
\usepackage{color}
\usepackage{epstopdf}
\usepackage{epsfig,float,afterpage,amssymb,wrapfig,psfrag}
\usepackage[caption=false]{subfig}

\usepackage{placeins}

\DeclareMathAlphabet{\bi}{OML}{cmm}{b}{it}

\begin{document}

\title{Global phase diagram of
a spin-orbit-coupled Kondo lattice model on the honeycomb lattice}
\author{Xin Li}
\affiliation{Beijing National Laboratory for Condensed Matter Physics and Institute of Physics, Chinese Academy of Sciences, Beijing 100190, China}
\affiliation{University of Chinese Academy of Sciences, Beijing 100049, China}
\author{Rong Yu}
\email{rong.yu@ruc.edu.cn}
\affiliation{Department of Physics and Beijing Key Laboratory of Opto-electronic Functional Materials and Micro-nano Devices, Renmin University of China, Beijing 100872, China
}
\author{Qimiao Si}
\email{qmsi@rice.edu}
\affiliation{Department of Physics \& Astronomy, Rice Center for Quantum Materials,
Rice University, Houston, Texas 77005,USA}
\date{\today}
\begin{abstract}
Motivated by the growing interest in the
novel quantum phases in materials with strong electron correlations and spin-orbit coupling,
we study the interplay
between the spin-orbit coupling, Kondo interaction, and magnetic frustration
of a Kondo lattice model
on a two-dimensional honeycomb lattice.
We
calculate the renormalized electronic structure and correlation functions
at the saddle point
based on a fermionic representation of the spin operators.
We find
a global phase diagram of the model at
half-filling, which
contains a variety of phases due to the competing interactions. In addition to a Kondo insulator,
there is a topological insulator with valence bond solid correlations in the spin sector, and two
antiferromagnetic phases.
Due to a competition between the spin-orbit coupling and Kondo interaction,
the direction of the magnetic moments
in the antiferromagnetic phases
can be either within
or perpendicular to
the lattice plane.
The latter antiferromagnetic state is topologically nontrivial for moderate and strong spin-orbit
couplings.
\end{abstract}


\maketitle

\section{Introduction}
Exploring novel quantum phases and the associated phase transitions in systems with strong electron correlations
is a major subject of contemporary condensed matter physics.\cite{SpecialIssue2010,Sachdev2011a,SiSteglich_Sci2010}
In this context, heavy fermion (HF) compounds play a crucial role. \cite{SiSteglich_Sci2010,GegenwartSi_NatPhys2007,Lohneysen_RMP2007,Tsunetsugu_RMP1997} In these materials,
the coexisted itinerant electrons and local magnetic moments (from localized $f$ electrons) interact via the antiferromagnetic
exchange coupling, resulting in the
 Kondo effect.\cite{Hewson_Book} Meanwhile, the Ruderman-Kittel-Kasuya-Yosida (RKKY) interaction, namely the exchange coupling among the local moments mediated by the itinerant electrons, competes with the Kondo effect.\cite{Doniach_Physica1977}
 This competition
 gives rise to
 a rich
 phase diagram with an antiferromagnetic (AFM) quantum critical point (QCP) and various emergent phases nearby.\cite{Custers_Nat2003,SiSteglich_Sci2010}

In the HF metals,
experiments~\cite{Schroder_Nat2000,Paschen_Nat2004}
have provide strong evidence for
 local quantum criticality,~\cite{Si_Nat2001,Coleman_JPCM2001}
 which is characterized by the beyond-Landau physics of Kondo destruction
 at the AFM QCP. Across this local QCP, the Fermi surface jumps from large in the paramagnetic HF liquid phase to small in the AFM
 phase with Kondo destruction.
 A natural question is how this local QCP
  connects
  to the conventional spin density wave (SDW) QCP, described by the Hertz-Millis theory~\cite{Hertz_1976,Millis_1993}.
A proposed global phase diagram
 \cite{Si_PhysB2006,Si_PSSB2010,Pixley_PRL2014,SiPaschen}
 makes this connection
   via the tuning
   of the quantum fluctuations
   in the local-moment magnetism.
  Besides the HF metals, it is also interesting to know whether a similar global phase diagram can be realized in Kondo insulators (KIs), where the chemical potential is inside the Kondo hybridization gap when the electron filling is commensurate.
  The KIs are nontrivial band insulators because the band gap originates from strong electron-correlation effects.
 A Kondo-destruction transition
 is expected to accompany the closure of the band gap.
 The question that remains open is
 whether the local moments immediately order or form a
 different type of magnetic states, such as spin liquid
 or
 valence bond solid (VBS),
  when the Kondo destruction takes place.

Recent years have seen extensive
studies about
the effect of a fine spin-orbit coupling (SOC) on the electronic bands.
In topological insulators (TIs), the bulk band gap opens due to a
nonzero
SOC,
and
there exist gapless surface states.
The
 nontrivial topology of the bandstructure is protected by the time reversal symmetry (TRS).
 Even for a system with broken TRS, the conservation of combination of TRS and translational symmetry can give rise to a topological antiferromagnetic insulator (T-AFMI).\cite{MongMoore_2010}
In general, these TIs and TAFIs can be tuned to topologically trivial insulators via  topological quantum phase transitions. But how the strong electron correlations influence the properties of these symmetry dictated topological phases and related phase transitions is still under active discussion.

The SOC also has important effects in
HF materials~\cite{SiPaschen}.
For example,
the SOC
can produce a topologically nontrivial bandstructure
and
induce exotic Kondo physics.\cite{Nakatsuji_PRL2006,Chen_PRB2017}
it may give
 rise to a topological Kondo insulator (TKI),\cite{Dzero_PRL2012}
which
has been invoked to understand
the resistivity plateau of the heavy-fermion SmB$_6$ at low temperatures.\cite{SmB6}.

From
a more general perspective,
SOC provides an additional tuning parameter enriching the global phase diagram of HF
systems~\cite{SiPaschen,YamamotoSi_JLTP2010}.
Whether and how the topological nontrivial quantum phases can emerge
in this phase diagram is
a timely issue.
Recent studies have advanced a Weyl-Kondo semimetal phase \cite{Lai2018}.
Experimental evidence has come from the new heavy fermion compound
Ce$_3$Bi$_4$Pd$_3$,  which display thermodynamic \cite{Dzsaber2017} and
zero-field Hall transport \cite{Dzsaber2018} properties that provide evidence for the salient features
of the Weyl-Kondo semimetal. These measurements respectively probe
a linearly dispersing electronic excitations with a velocity that is renormalized by several
orders of magnitude and singularities in the Berry-curvature distribution.

This type of theoretical studies are also of interest for
a Kondo lattice model
defined on
a honeycomb lattice,\cite{Feng_PRL2012}
which
readily accommodates
the
SOC \cite{KaneMele_PRL2005} .
In the  dilute-carrier
 limit, this model supports a nontrivial Dirac-Kondo semimetal (DKSM) phase,
which can be tuned to a TKI by increasing SOC.\cite{Feng_2016}
 In Ref.~\onlinecite{Feng_PRL2012}, it was shown that, at half-filling,
 increasing the Kondo coupling induces a direct transition from a TI to a KI.
A related model, with the conduction-electron part of the Hamiltonian described by
a Haldane model~\cite{Haldane1988}
on the honeycomb lattice, was subsequently studied.\cite{Zhong_PRB2012}

Here we investigate the global phase diagram of a
spin-orbit-coupled Kondo lattice model on the honeycomb lattice
at half-filling. We show that the
competing interactions
 in this model give rise to a very rich phase diagram containing a TI, a KI, and two AFM phases.
 We focus on discussing the influence of magnetic frustration
 on the phase diagram.
 In the TI, the local moments develop a VBS order. In the two AFM phases, the moments are ordered, respectively, in the plane of the honeycomb lattice (denoted as AFM$_{xy}$) and perpendicular to the plane (AFM$_z$).
 Particularly in the AFM$_z$ phase, the conduction electrons may have a topologically nontrivial bandstructure,
 although the TRS is explicitly broken. This T-AFM$_z$ state connects to the trivial AFM$_z$ phase via a topological phase transition
 as
 the SOC
 is reduced.

 The remainder of the
 paper is organized as follows. We start by introducing the model and our theoretical procedure in Sec.II.
 In Sec.III we discuss the magnetic phase diagram of the Heisenberg model for the local moments.
 Next we obtain the global phase diagram of the full model in Sec. IV.
 In Sec V we examine the nature of the conduction-electron bandstructures in the AFM states,
 with a focus on their
 topological characters. We discuss the
 implications of our results
  in Sec. VI.

\section{Model and method}
The model we considere here is defined on an effective double-layer honeycomb lattice. The top layer contains conduction electrons realizing the Kane-Mele Hamiltonian~\cite{KaneMele_PRL2005}. The conduction electrons are Kondo coupled to (\emph{i.e.},
experiencing an AF exchange coupling $J_{\rm{K}}$ with) the localized magnetic
moments in the bottom layer. The local moments interact among themselves through direct exchange interaction
as well as the conduction electron mediated RKKY interaction;
this interaction is described
  by a simple $J_{1}$-$J_{2}$ model. Both the conduction bands and the localized bands are half-filled.
This Kondo-lattice
Hamiltonian takes the following form on the honeycomb lattice:
\begin{eqnarray}\label{Kondo}
H&=&t\sum_{\langle i j\rangle\sigma}c^{\dagger}_{i\sigma}
c_{j\sigma} + i\lambda_{\rm{so}}\sum_{\ll
 ij\gg\sigma\sigma'}v_{ij}c^{\dagger}_{i\sigma}
{\sigma}^z_{\sigma\sigma'}
c_{j\sigma'}\nonumber\\&+&J_K\sum_i {\vec s}_i\cdot {\vec
S}_i+J_{1}\sum_{\langle i j\rangle}{\vec S}_i\cdot{\vec S}_j
+J_{2}\sum_{\langle \langle i j\rangle \rangle}{\vec S}_i\cdot{\vec S}_j,
\end{eqnarray}
where $c^\dagger_{i\sigma}$ creates a conduction electron at site $i$ with spin index $\sigma$. $t$ is the hopping parameter between the nearest neighboring (NN) sites, and $\lambda_{\rm{so}}$ is the strength of the SOC between next-nearest neighboring (NNN) sites. $v_{ij}=\pm1$, depending on the direction of the NNN hopping. ${\vec s}_i=c^{\dagger}_{i\sigma} \vec{\sigma}_{\sigma\sigma'}c_{i\sigma'}\nonumber$, is the spin operator of the conduction electrons at site $i$ with $\vec{\sigma}=\sigma^{x} , \sigma^{y} , \sigma^{z}$ being the pauli matrices. ${\vec S}_i$ refers to the spin operator of the local moments with spin size $S=1/2$. In the model we considered here, $J_{\rm{K}}$, $J_1$, and $J_2$ are all AF.
By incorporating the Heisenberg
interactions, the Kondo-lattice model we study readily captures the effect of geometrical frustration.
 In addition, instead of treating the
  Kondo screening and magnetic order in terms of the longitudinal and transverse components of the Kondo-exchange interactions \cite{Lacroix_prb1979,GMZhang,Zhong_PRB2012}, we will treat both effects in terms of
interactions that are spin-rotationally invariant;
this will turn out to be important in mapping out the global phase diagram.

\begin{figure}[!th]
  \centering
\includegraphics[width=0.8\columnwidth]{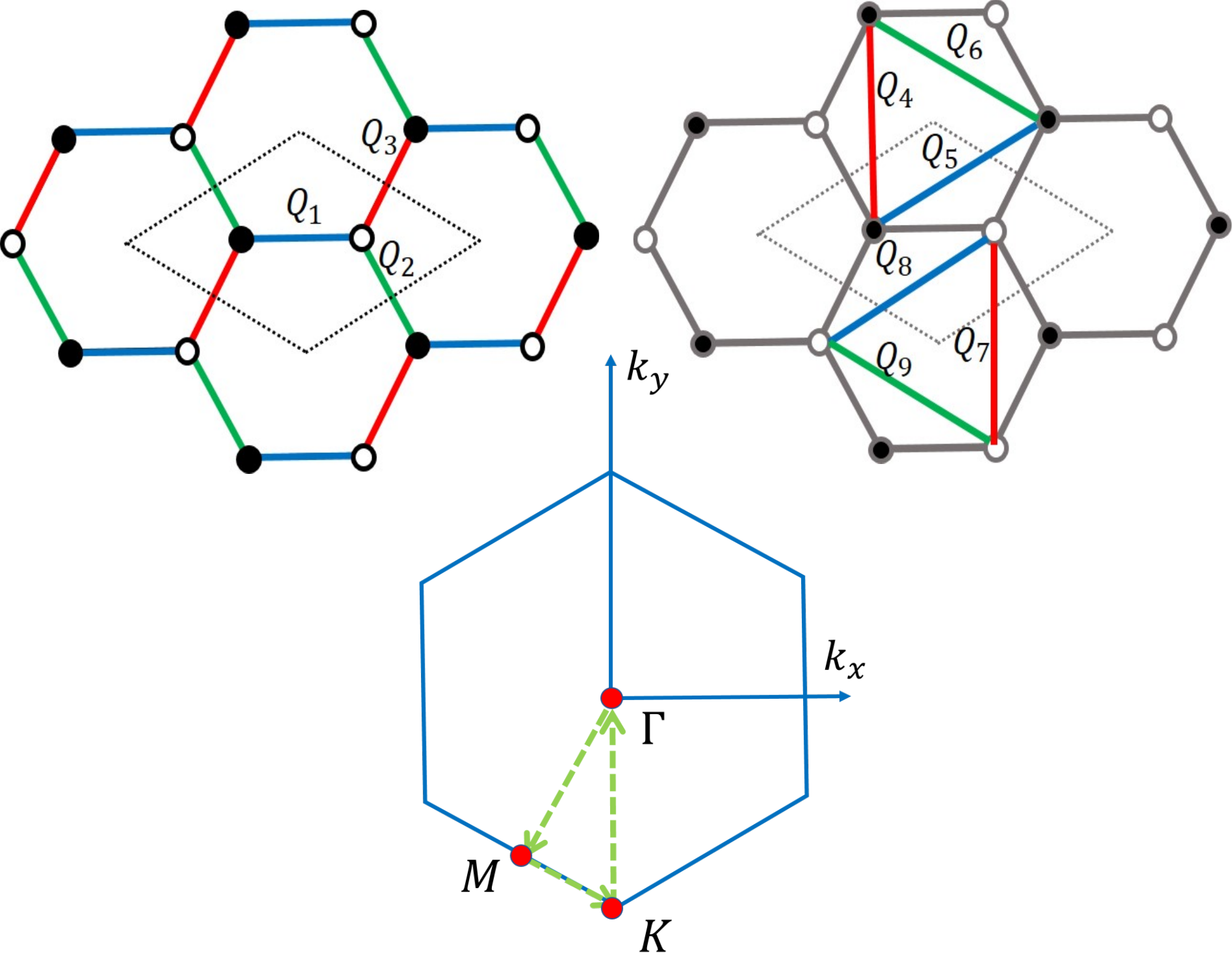}
 \caption{Top panels: Definition of nearest neighboring and next nearest neighboring valence bond
 fields $Q_{ij}$. Filled and empty
 circles denote
  the two sublattices A and B, respectively.
 Different bond directions are labeled by different colors. Bottom panel:
 First Brillouin zone
 corresponding
 to the two-sublattice unit cell.}
\label{fig1}
\end{figure}

We use the spinon representation for ${\vec S}_i$, \emph{i.e.}, by rewriting ${\vec S}_i=f^{\dagger}_{i\sigma} \vec{\sigma}_{\sigma\sigma'}f_{i\sigma'}\nonumber$ along with the constraint $\sum_{\sigma} f^{\dagger}_{i\sigma} f_{i\sigma}=1$, where $f^\dagger_{i\sigma}$ is the spinon operator. The constraint is enforced by introducing the Lagrange multiplier term $\sum_{i} \lambda_{i} (\sum_{\sigma} f^{\dagger}_{i\sigma} f_{i\sigma} -1) $ in the Hamiltonian. In order to
 study both the non-magnetic and magnetic phases,
we
 decouple
 the Heisenberg Hamiltonian into two channels:
\begin{eqnarray}\label{Eq:HeisMF}
&&J\bm{ S}_{i} \cdot \bm{S}_{j} \nonumber  \\
&=&x J \bm{ S}_{i} \cdot \bm{S}_{j} +(1-x) J \bm{ S}_{i} \cdot \bm{S}_{j}  \nonumber \\
&\simeq & x \left ( \frac{J}{2} |Q_{ij} |^{2} -\frac{J}{2}  Q^{*}_{ij} f_{i\alpha}^{\dagger} f_{j\alpha} -\frac{J}{2}  Q_{ij} f_{j\alpha}^{\dagger} f_{i\alpha} \right)  \nonumber \\
&+& (1-x) \left( -J \bm{M}_{i} \cdot \bm{M}_{j}+ J \bm{M}_{j} \cdot \bm{S} _{i} + J \bm{M}_{i} \cdot \bm{S} _{j}  \right)
\end{eqnarray}
Here $x$ is a
parameter that
is introduced in keeping with the generalized procedure of Hubbard-Stratonovich decouplings
and will be fixed
 to conveniently describe the effect of quantum fluctuations.
The corresponding
valence bond (VB) parameter $Q_{ij}$ and sublattice magnetization $\bm{M}_i$ are $Q_{ij}=\langle \sum_{\alpha} f_{i\alpha}^{\dagger} f_{j \alpha} \rangle$ and $\bm{M}_i=\langle \bm{S}_{i}\rangle$, respectively.
Throughout this paper, we consider the two-site unit cell thus excluding any states that breaks lattice translation symmetry. Under this construction, there are 3 independent VB mean fields $Q_{i}$, $i=1,2,3$, for the NN bonds and 6 independent VB mean fields $Q_{i}$, $i=4,5,...,9$, for the NNN bonds.
They are illustrated in Fig. \ref{fig1}.
We consider only AF exchange interactions,
$J_{1}>0$ and $J_{2}>0$,
and
will
thus
only take into account AF order
with $\bm{M}=\bm{M}_{i\in A}=-\bm{M}_{i \in B}$.

\begin{figure}[!th]
  \centering
  \includegraphics[width=0.8\columnwidth]{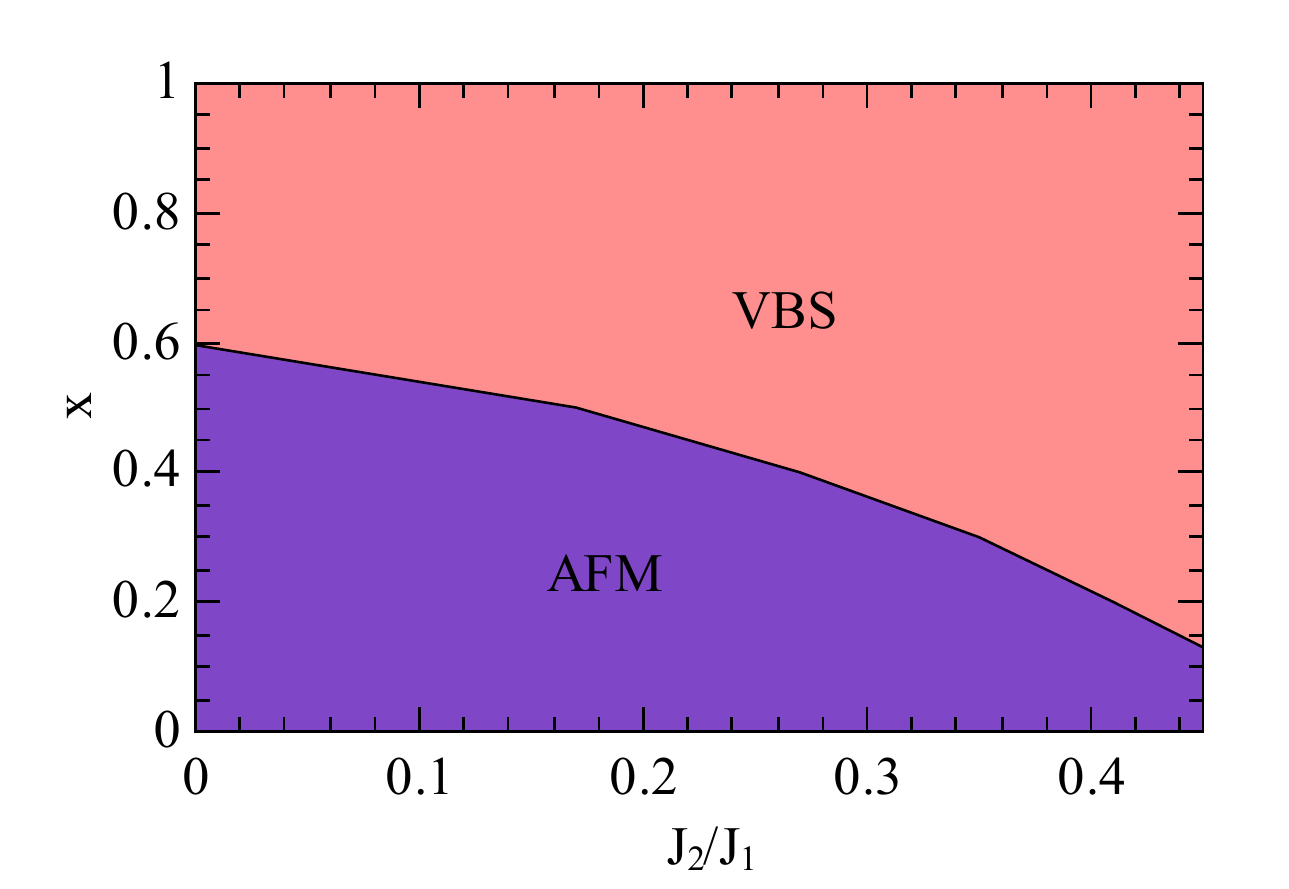}
 \caption{Ground-state phase diagram of the $J_1$-$J_2$ Hamiltonian for the local moments in the $x$-$J_{2}/J_{1}$ plane. A NN VBS and an AFM state are stabilized in the parameter regime shown.}
\label{fig2}
\end{figure}

To take into account the Kondo hybridization and the possible magnetic order on an equal footing,
we follow the treatment of the Heisenberg interaction as outlined in Eq.~\ref{Eq:HeisMF} and
decouple the Kondo interaction as follows:
\begin{eqnarray}\label{Eq:JKMF}
&& J_{K} \bm{S} \cdot  \bm{s}  \nonumber  \\
&\simeq & y  \left(\frac{J_{K} }{2}  |b|^{2} -\frac{J_{K} }{2} b f^{\dagger}_{i\alpha} c_{i \alpha}
-\frac{J_{K} }{2} b^{*} c_{i \alpha}^{\dagger}  f_{i\alpha} \right)  \nonumber   \\
&+& (1-y)  \left( - J_{K} \bm{M}_{i} \cdot \bm{m}_{i} + J_{K} \bm{S}_{i}  \cdot \bm{m}_{i} +  J_{K} \bm{s}_{i}  \cdot \bm{M}_{i}   \right).
\end{eqnarray}
Here we have introduced the mean-field parameter for the Kondo hybridization, $b=\langle \sum_{\alpha} c_{i\alpha}^{\dagger} f_{i\alpha} \rangle $, and the conduction electron magnetization: $\bm{m_{i}} = \langle \bm{s}_{i} \rangle$. For nonzero $b$, the conduction band will Kondo hybridize with the local moments and the system at half-filling is a KI. On the other hand, when $b$ is zero and $\bm{M}$ is nozero, magnetization ($\bm{m}\neq0$) on the conduction electron band will be induced by the Kondo coupling, and various AF orders can be stabilized depending on the strength of the SOC.
Just like the parameter $x$ of Eq.~\ref{Eq:HeisMF} is chosen
so that a saddle-point treatment captures the
quantum fluctuations in the form of spin-singlet bond parameters~\cite{Pixley_PRL2014},
the parameter $y$ will be specified according to the criterion that the treatment at the same level
describes the quantum fluctuations in the form of Kondo-insulator state (see below).

\begin{figure}[th!]
  \centering
  \includegraphics[width=0.8\columnwidth]{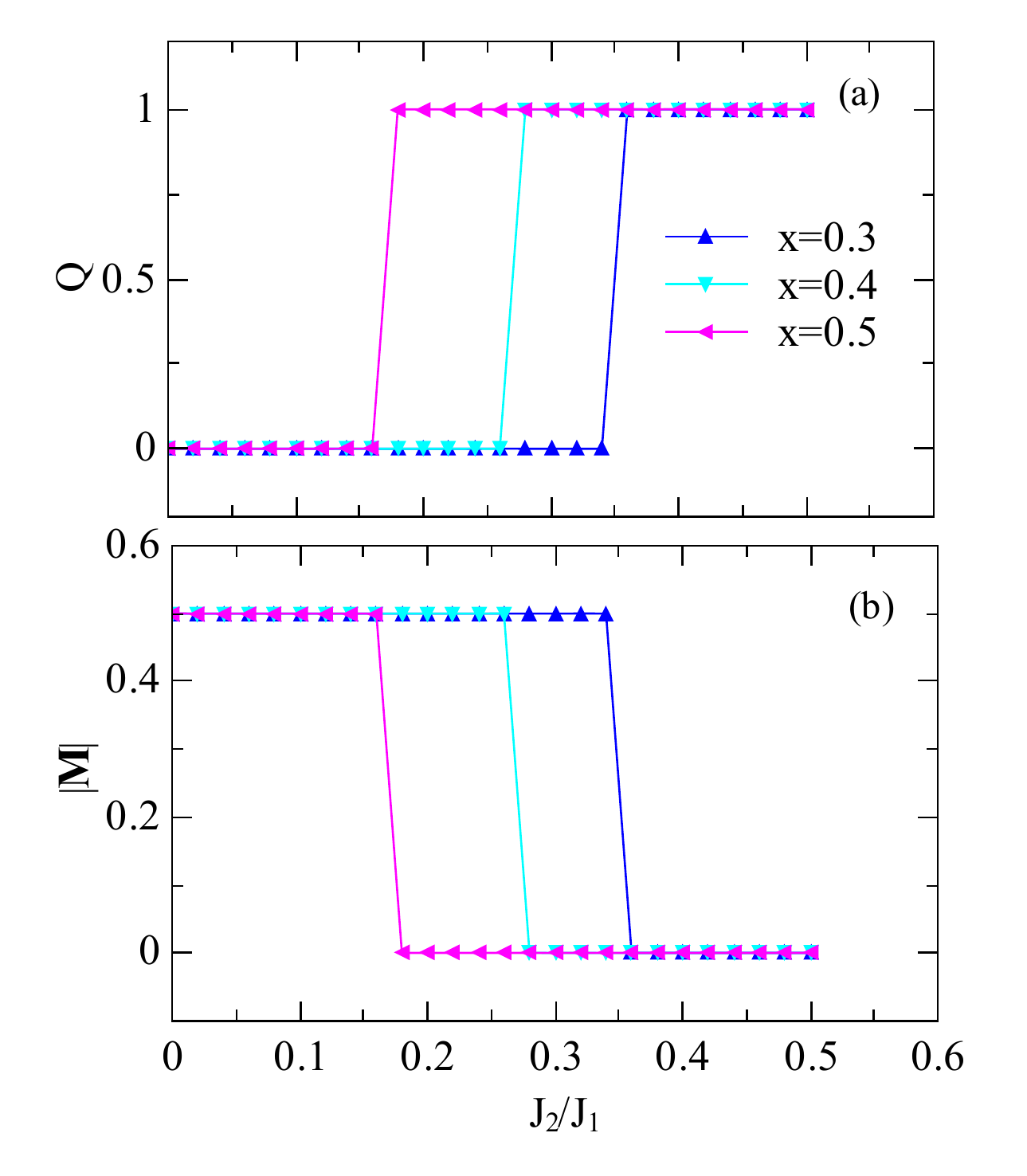}
 \caption{Evolution of the VBS order parameter $Q$ [in (a)] and the AFM order parameter $M$ [in (b)] as a function of $J_{2}/J_{1}$ for $x=0.3,0.4,0.5$.}
\label{fig3}
\end{figure}

\section{Phase diagram of the Heisenberg model for the local moments}
Because of the complexity of the full Hamiltonian,
we  start by setting $J_{K}=0$ and discuss the possible ground-state phases
of the $J_1$-$J_2$ Heisenberg model
for the local moments. By
treating the problem at the saddle-point level
in Eq.~\eqref{Eq:HeisMF}, we obtain the phase diagram in the $x$-$J_{2}/J_{1}$ plane shown in Fig.\ref{fig2}.
Here the $x$-dependence is studied in the same spirit as that of Ref.~\onlinecite{Pixley_PRL2014}
for the Shastry-Sutherland lattice.
In the parameter regime explored, an AF ordered phase (labeled as ``AFM" in the figure)
and a valence bond solid (VBS) phase are stabilized. The AF order stabilized is the two-sublattice N\'{e}el order
on the honeycomb lattice, and the VBS order refers to covering of dimer singlets with $|Q_{i}|=Q \neq 0 $ for one out of the three NN bonds (e.g. $Q_{1}\neq 0, Q_{2}=Q_{3}=0$) and $|Q_{i}|=0$ for all the NNN bonds.
This VBS state spontaneously breaks the C$_{3}$ rotational symmetry of the lattice.
We thus define the order parameter for VBS state to be $Q=|\sum_{j=1,2,3} Q_{j} e^{ i  (2\pi j/3) } |$.

In Fig.~\ref{fig3} we plot the evolution of VBS and AF order parameters $Q$ and $M$
as a function of $J_{2}/J_{1}$. A direct first-order transition
(signaled by the mid-point of the jump of the order parameters) between these two phases is observed
for $ x \lesssim 0.6$.
For
 the sake of understanding the global phase diagram of the full Kondo-Heisenberg model,
 we limit our discussion to $J_2/J_1<1$, where only the NN VBS is relevant.
A different decoupling scheme approach was used to study
this model~\cite{Liu_JPCM2016} found results that are, in the parameter regime of overlap,
 consistent with ours.
To fix the parameter $x$,
we compare our results with
those about the $J_1-J_2$ model
derived from previous numerical studies.
DMRG studies~\cite{Ganesh_PRL2013} found
that
 the AFM state is stabilized for $J_{2}/J_{1}<0.22$, and VBS exists for $J_{2}/J_{1}>0.35$, while in between the nature of the ground states are still under debate. In this parameter regime,
 the DMRG
 calculations suggest
 a plaquette resonating valence bond (RVB) state,\cite{Ganesh_PRL2013} while other methods
 implicate possibly spin liquids.\cite{Clark_PRL2011} In light of these numerical results, we take $x=0.4$ in our
 calculations.
 This leads to a direct transition from AFM to VBS at $J_{2}/J_{1} \simeq 0.27$, close to the values of phase boundaries of these two phases determined by other numerical methods.

\section{Global phase diagram of the
Kondo-lattice model
}
\begin{figure}[!th]
  \centering
  \includegraphics[width=0.8\columnwidth]{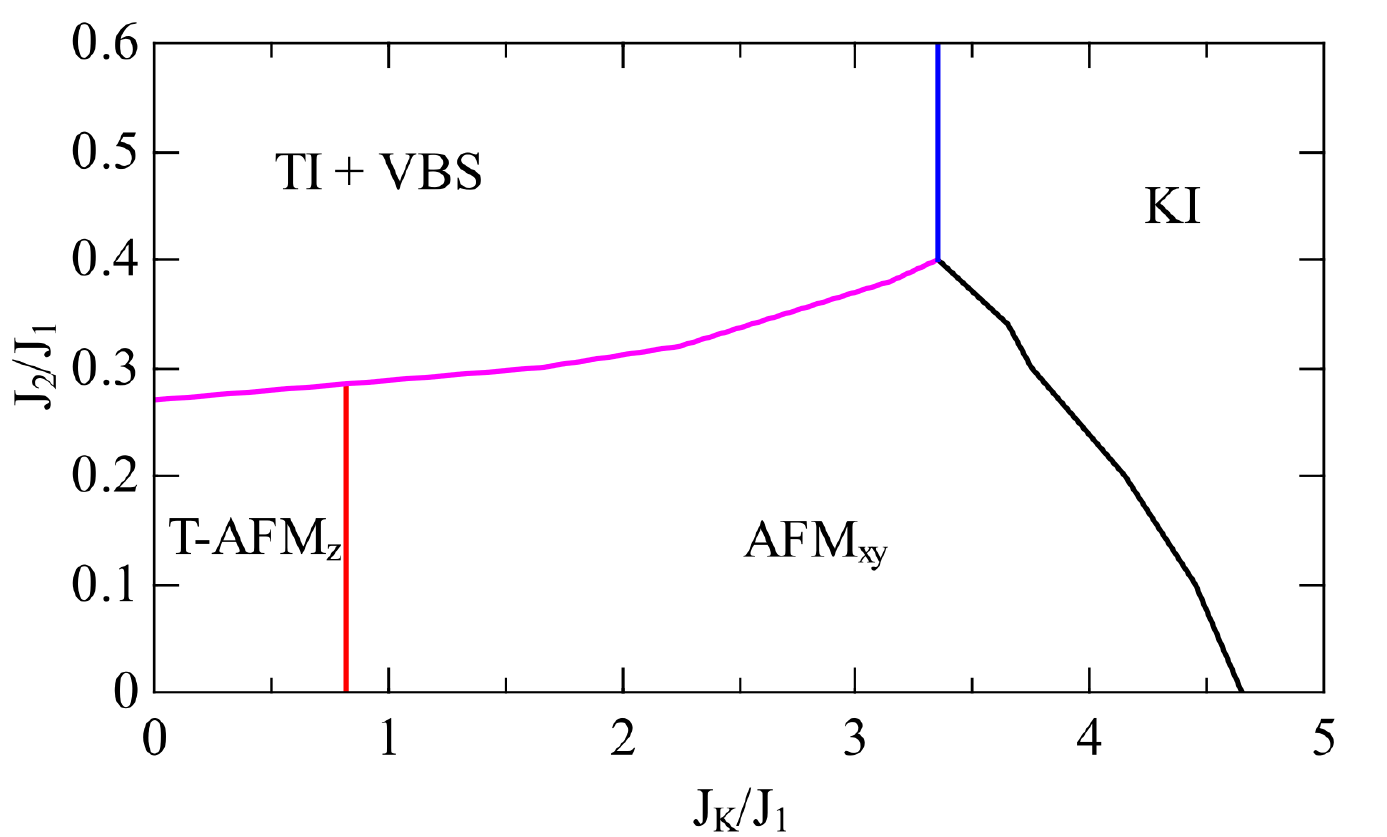}
 \caption{Global phase diagram at $T=0$
 from the saddle-point calculations with
$x=0.4$, $y=0.7$.
 The ground states include the valence-bond solid (VBS) and Kondo insulator (KI),
 as well as two antiferromagnetic orders,
 T-AFM$_z$ and AFM$_{xy}$, as described in Sec.~V.
 }
\label{fig4}
\end{figure}

We now turn to
 the global phase diagram of the full model by turning on the Kondo coupling.
 For definiteness, we set $J_{1}=1$ and
consider $t=1$ and $\lambda_{so}=0.4$.
As prescribed in the previous section, we take $x=0.4$.
Similar considerations for $y$ require that its value allows for quantum fluctuations in the form of
Kondo-singlet formation. This has guided us to take
$y=0.7$
(see below). The corresponding phase diagram as a function of
$J_{K}$ and the frustration parameter $J_2/J_1$
 is shown in Fig.~\ref{fig4}.

\begin{figure}[!th]
  \centering
  \includegraphics[width=0.8\columnwidth]{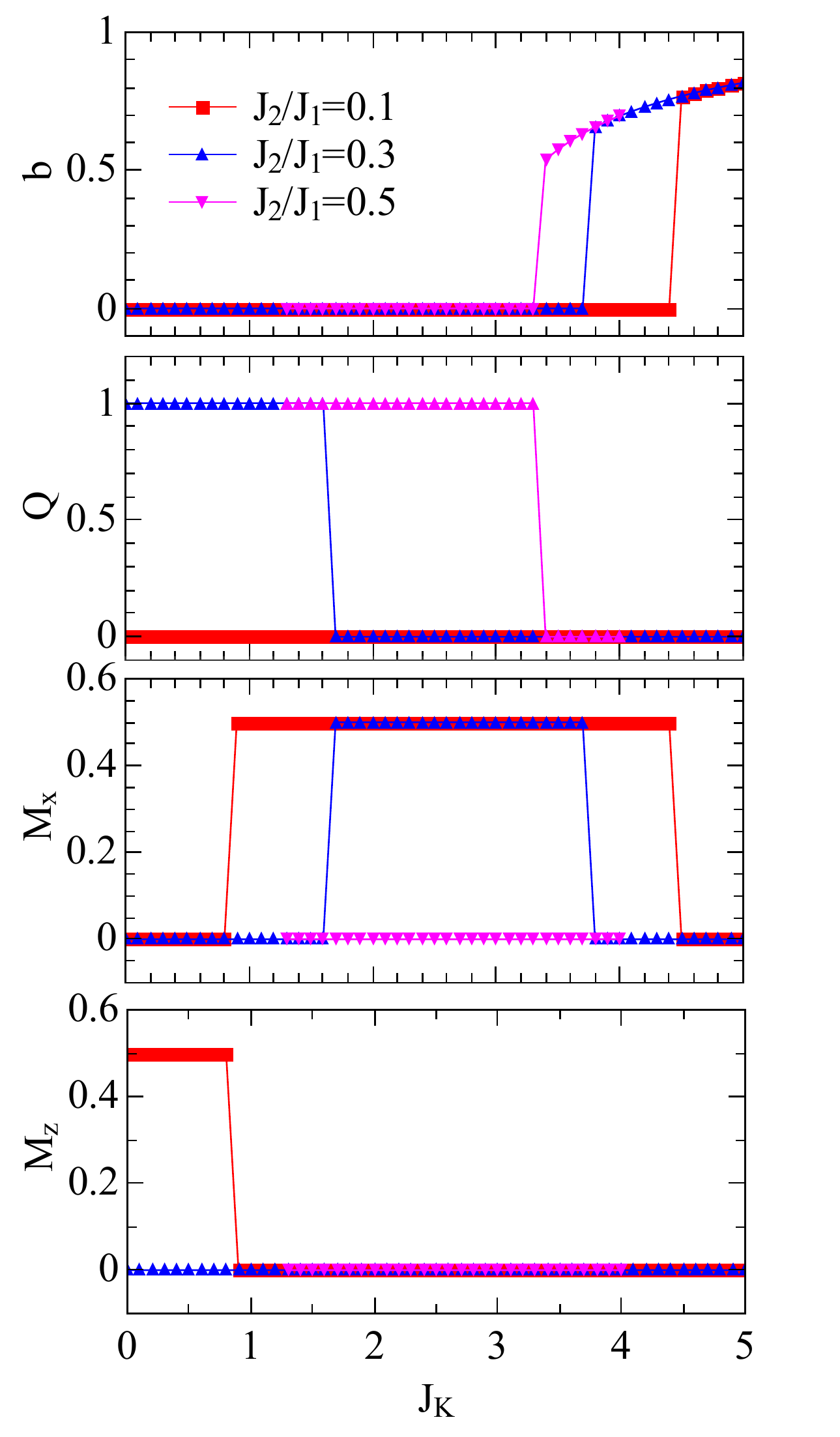}
 \caption{Evolution of the
parameters $b$  , $Q$  , $M_{x}$ and $M_{z}$  as a function of $J_{K}$ for different ratio of $J_{2}/J_{1}$. }
\label{fig5}
\end{figure}

In our calculation, the phase boundaries are determined by sweeping $J_{K}$ while along multiple horizontal cuts for several fixed $J_{2}/J_{1}$ values, as shown in Fig. 5. For small $J_K$ and large $J_2/J_1$, the local moments and the conduction electrons are still effectively decoupled. The conduction electrons form a TI for finite SOC, and the local moments are in the VBS ground state as discussed in the previous section. When both $J_K$ and $J_2/J_1$ are small, the ground state is AFM. Due to the Kondo coupling, finite magnetization $\bm{m}$ is induced for the conduction electrons. This opens
a
spin density wave (SDW) gap in the conduction band, and therefore the ground state of the system is an AFM insulator. The SOC couples the rotational symmetry in the spin space to the one in real space. As a consequence, the ordered moments in the AFM phase can be either along the $z$ direction (AFM$_{z}$) or in the $x$-$y$ plane (AFM$_{xy}$). For finite SOC, these two AFM states have different energies, which can be tuned by $J_K$. As shown in the phase diagram, the AFM phase contains two ordered states, the AFM$_{z}$ and AFM$_{xy}$. They are separated by a spin reorientation transition at $J_K/J_1\approx0.8$. For the value of SOC taken, the AFM state is topologically nontrivial, and is hence denoted as T-AFM$_z$ state. The nature of this state and the associated topological phase transition is discussed in detail in the next section.

For sufficiently large $J_K$, the Kondo hybridization $b$ is nonzero (see Fig.5(a)), and the ground state is a KI. Note that for finite SOC, this KI does not have a topological nontrivial edge state, as a consequence of the topological no-go theorem~\cite{Feng_PRL2012,HasanKane_RMP2010,QiZhang_RMP2011}. In our calculation
at the saddle-point level, the KI exists
for $y \geq 0.6$;
this provides the basis for taking $y=0.7$, as noted earlier.
Going beyond the saddle-point level,
the dynamical effects of the Kondo coupling will appear,
and we will expect the KI phase to arise for other choices of $y$.


Several remarks are in order. The phase diagram, Fig.~\ref{fig4}, has a similar profile of the global phase diagram for the Kondo insulating systems~\cite{YamamotoSi_JLTP2010,Pixley_PRB2018}.
However, the presence
of SOC has enriched the phase diagram.
In the AF state, the ordered moment may lie either within the plane or be perpendicular to it.
These two states have very different topological properties.
We now turn to a detailed discussion of this last point.

\section{
Topological properties of
 the AFM states}
\begin{figure}[!th]
  \centering
  \includegraphics[width=0.8\columnwidth]{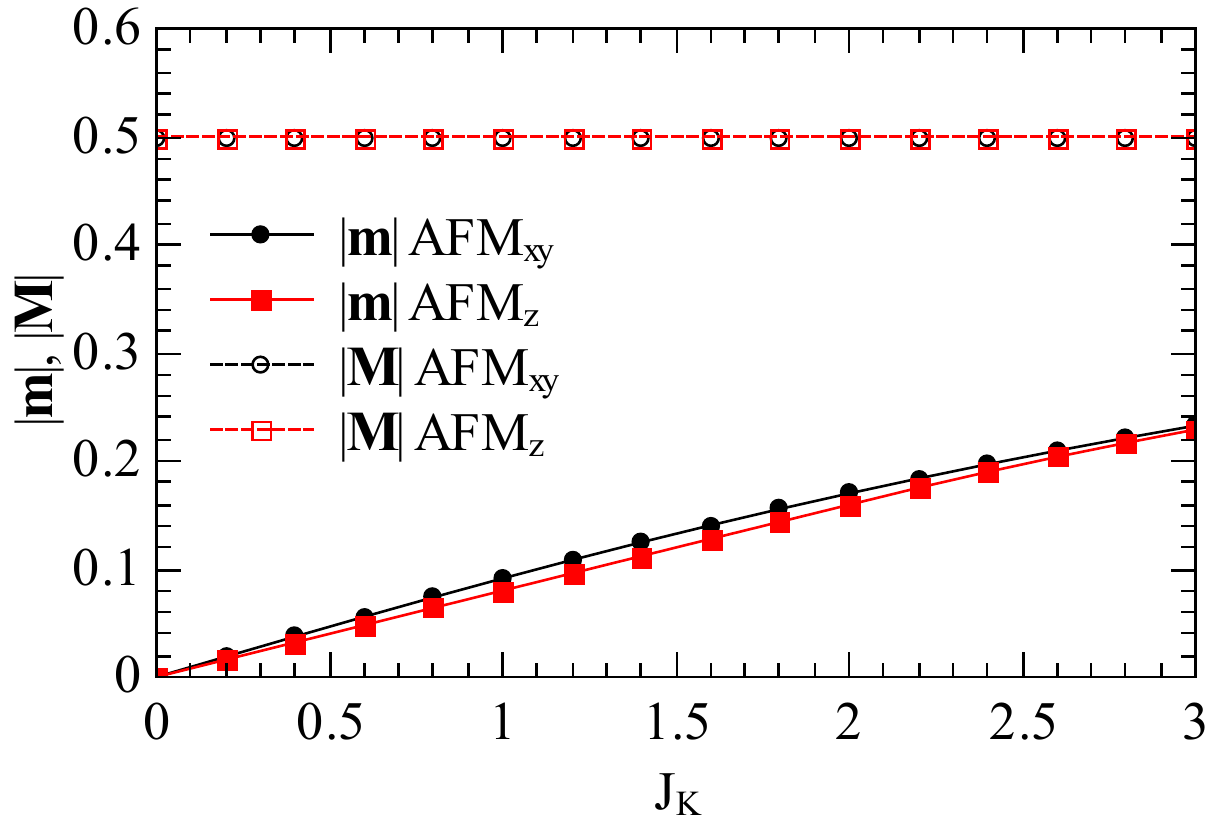}
 \caption{
 The
  conduction electron magnetization
   for the AFM$_{xy}$ and AFM$_{z}$ states at $\lambda_{so}=0.1$.}
\label{fig6}
\end{figure}

In this section we discuss the properties of the AFM$_{xy}$ and AFM$_{z}$ states, in particular to address their topological nature. For a clear discussion, we fix $t=1$, $J_1=1$, and $J_2$=0. Since the Kondo hybridization is
not essential to the nature of
the AFM states,
in this section we simply the discussion by setting
$y=0$.

We start by defining the order parameters of the two states:
\begin{eqnarray}
M_{x}&=&\langle S_{f,A}^{x} \rangle= -\langle S_{f,B}^{x} \rangle , \\
M_{z}&=&\langle S_{f,A}^{z} \rangle= - \langle S_{f,B}^{z} \rangle, \\
m_{x}&=& -\langle s_{c,A}^{x} \rangle= \langle s_{c,B}^{x} \rangle, \\
m_{z}&=& -\langle s_{c,A}^{z} \rangle= \langle s_{c,B}^{z} \rangle.
\end{eqnarray}
Note that for AFM$_{xy}$ state we set $M_x=m_{y}= 0$ without losing generality.
In Fig.(\ref{fig6}) we plot the evolution of these AFM order parameters with $J_K$ for a representative value of SOC $\lambda_{so}=0.1$. Due to the large $J_1$ value we take, the sublattice magnetizations of the local moments are already saturated to $0.5$. Therefore,
at the saddle-point level,
 they serve as effective (staggered) magnetic fields to the conduction electrons. The Kondo coupling then induces finite sublattice magnetizations for the conduction electrons, and they increase linearly with $J_K$ for small $J_K$ values. But $m_x$ is generically different from $m_z$. This is important for the stabilization of the states.

We then discuss the energy competition between the AFM$_{xy}$ and AFM$_{z}$ states.
The conduction electron part of the mean-field Hamiltonian reads:
\begin{equation}
H_{c}=
\begin{pmatrix}
c_{A\uparrow}^{\dagger} &
c_{A\downarrow}^{\dagger}&
c_{B\uparrow}^{\dagger} &
c_{B\downarrow}^{\dagger} &
\end{pmatrix}^{T}
h_{MF}
\begin{pmatrix}
c_{A\uparrow}  \\
c_{A\downarrow}  \\
c_{B\uparrow}  \\
c_{B\downarrow}  \\
\end{pmatrix}
\end{equation}

with
\begin{equation}
\label{afmx}
h_{MF}=
\begin{pmatrix}
\Lambda(k) &  J_{K}M_{x}/2  & \epsilon(k) & \\
 J_{K}M_{x}/2 &  -\Lambda(k) &  & \epsilon(k) \\
\epsilon^{*}(k) &   & -\Lambda(k) & -J_{K}M_{x}/2 \\
 &  \epsilon^{*}(k) & -J_{K}M_{x}/2  &   \Lambda(k)
\end{pmatrix}
\end{equation}

for the AFM$_{xy}$ state and

\begin{widetext}
\begin{equation}
\label{afmz}
h_{MF}=
\begin{pmatrix}
\Lambda(k)+J_{K}M_{z}/2 &   & \epsilon(k) & \\
 &  -\Lambda(k)-J_{K}M_{z}/2 &  & \epsilon(k) \\
\epsilon^{*}(k) &   & -\Lambda(k)-J_{K}M_{z}/2 &  \\
 &  \epsilon^{*}(k) &   &   \Lambda(k)+J_{K}M_{z}/2
\end{pmatrix}
\end{equation}
\end{widetext}
for the AFM$_{z}$ state. Here $\Lambda(k)= 2 \lambda_{so} \left(  sin( k \cdot a_{1}) - sin( k \cdot a_{2}) - sin( k \cdot (a_{1}-a_{2})  ) \right) $,
$\epsilon(k)=t_{1} ( 1+ e^{-i k \cdot a_{1} } + e^{-i k \cdot a_{2}} ) $, $\epsilon^{*}(k)$ is the complex conjugate of $\epsilon(k)$, and $a_{1}=(\sqrt{3}/2, {1}/{2})$,$a_{2}=(\sqrt{3}/2, -{1}/{2})$ are the primitive vectors.
For both states the eigenvalues are doubly degenerate.
\begin{eqnarray}
E^{c}_{\pm,xy}(k)&=&\pm \sqrt{ \Lambda(k)^{2}+(J_{K}M_{x}/2)^{2} + |\epsilon(k)|^{2} }  \label{Eq:Ecxy}\\
E^{c}_{\pm,z}(k)&=&\pm \sqrt{ (\Lambda(k)+J_{K}M_{z}/2)^{2} + |\epsilon(k)|^{2} } \label{Eq:Ecz}
\end{eqnarray}

The eigenenergies of the spinon band can be obtained in a similar way:
\begin{eqnarray}
E^{f}_{\pm,xy}(k)& = &\pm \frac{1}{2} (3 J_{1} M_{x} + J_{K} m_{x} ), \label{Eq:Efxy}\\
E^{f}_{\pm,z}(k) & = &\pm \frac{1}{2} (3 J_{1} M_{z} + J_{K} m_{z} ). \label{Eq:Efz}
\end{eqnarray}
The expression of total energy for either state is then
\begin{eqnarray}
E_{tot}
&=&2 \frac{1}{N_{k}} \sum_{k}E^{c}_{-}(k)
+2 \frac{1}{N_{k}} \sum_{k}E^{f}_{-}(k) \nonumber \\
&+&
3J_{1} |\bm{M}|^{2}
+2J_{K} (\bm{M}\cdot\bm{m}).\label{Eq:Etot}
\end{eqnarray}
The first line of the above expression comes from filling the bands up to the Fermi energy (which is fixed to be zero here). The second line is the constant term in the mean-field decomposition. The factor of $2$ in the $k$ summation is to take into account the double degeneracy of the energies. $N_{k}$ refers to the number of $k$ points in the first Brillouin zone.

By comparing the expressions of $E_{-}^{c}(k)$ in Eqns.~\eqref{Eq:Ecxy} and \eqref{Eq:Ecz}, we find that adding a small $M_{x}$ is to increase the size of the gap at both of the two (inequivalent) Dirac points, thereby pushing the states further away from the Fermi-energy. While adding a small $M_{z}$ is to enlarge the gap at one Dirac point but reduce the gap size at the other one. Therefore, an AFM$_{xy}$ state is more favorable than the AFM$_z$ state in lowering the energy of conduction electrons $\sum_{k} E_{-}^{c}(k)$.

On the other hand, from Eqns.\eqref{Eq:Efxy}-\eqref{Eq:Etot}, we see that the overall effect of adding a magnetization of the conduction band, $\bm{m}$, is to increase the total energy $E_{tot}$ (the main energy increase comes from the $2J_{K} (\bm{M}\cdot\bm{m})$ term). Because $|m_z|<|m_x|$ from the self consistent solution, as shown in Fig.~\ref{fig6}, the energy increase of the AFM$_{z}$ state is smaller than that in the AFM$_{xy}$ state.

\begin{figure}[!th]
  \centering
  \includegraphics[width=0.8\columnwidth]{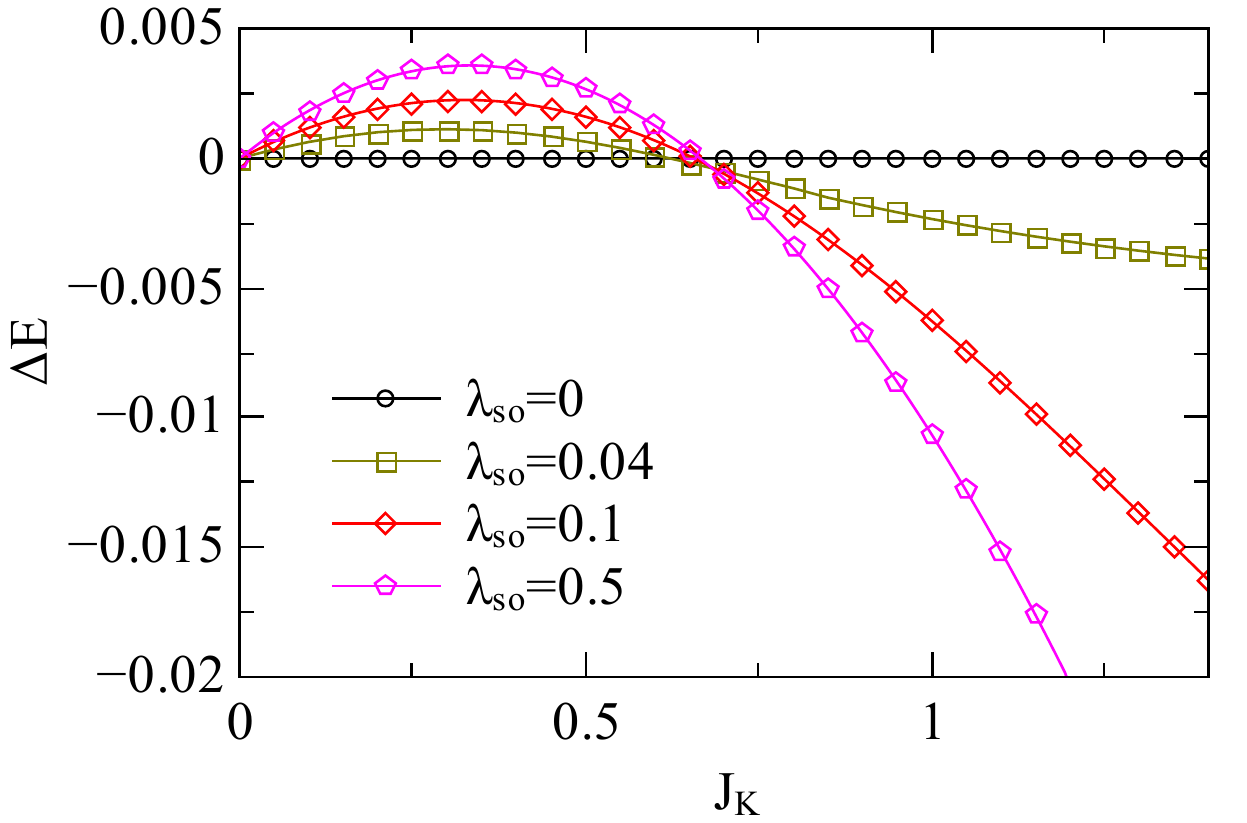}
 \caption{
 The energy
  difference between AFM$_{z}$ and AFM$_{xy}$
  states
  as a function of $J_{K}$ for various
  values
  of spin-orbital coupling $\lambda_{so}$. }
\label{fig7}
\end{figure}

With increasing $J_K$ the above two effects from the magnetic orders compete, resulting in different magnetic ground states as shown in Fig.~\ref{fig4}. This analysis is further supported by our self-consistent mean-field calculation. In Fig.~\ref{fig7} we plot the energy difference between these two states $\Delta E = E_{xy}-E_{z}$ as a function of $J_{K}$ at several $\lambda_{so}$ values. In the absence of SOC, the model has the spin SU(2) symmetry, and the AFM$_z$ and AFM$_{xy}$ states are degenerate with $\Delta E=0$. For finite $\lambda_{so}$, at small $J_K$ values, the energy gain from the $\sum_{k} E_{-}^{c}(k)$ term dominates, $\Delta E>0$, and the ground state is an AFM$_{z}$ state. With increasing $J_K$, the contribution from the $2J_{K} (\bm{M}\cdot\bm{m})$ term is more important. $\Delta E$ crosses zero to be negative, and the AFM$_{xy}$ state is eventually energetically favorable for large $J_K$.


Next we discuss the topological nature of the AFM$_z$ and AFM$_{xy}$ state. In the absence of Kondo coupling $J_K$, the conduction electrons form a TI, which is protected by the TRS. Their the left- and right-moving edge states connecting the conduction and valence bands are respectively coupled to up and down spin flavors (eigenstates of the $S^z$ operator) as the consequence of SOC, and these two spin polarized edge states do not mix.

Once the TRS is broken by the AFM order, generically, topologically nontrivial edge states are no longer guaranteed. However, in the AFM$_z$ state, the structure of the Hamiltonian for the conduction electrons is as same as that in a TI. This is clearly shown in Eq.~\eqref{afmz}: the effect of magnetic order is only to shift $\Lambda(k)$ to $\Lambda(k)+J_{K}M_{z}/2$. In particular, the spin-up and spin-down sectors still do not mix each other. Therefore, the two spin polarized edge states are still well defined as in the TI, and the system is topologically nontrivial though without the protection of TRS.
Note that the above analysis is based on assuming $J_K M_z\ll\Lambda(k)$, where the bulk gap between the conduction and valence bands is finite. For $J_K M_z>6\sqrt{3}\lambda_{so}/(1-y)$, the bulk gap closes at one of the inequivalent Dirac points and the system is driven to a topologically trivial phase via a topological phase transition.\cite{Feng_PRL2012}.
 We also note that a similar AFM$_z$ state arises in a Kondo lattice model without SOC but with
 a Haldane coupling,
 as analyzed in Ref.~\onlinecite{Zhong_PRB2012}.

For the AFM$_{xy}$ state, we can examine the Hamiltonian for the conduction electrons in a similar way. As shown in Eq.~\eqref{afmx}, the transverse magnetic order $M_x$ mixes the spin-up and spin-down sectors. As a result, a finite hybridization gap opens between the two edge states making the system topologically trivial.

\begin{figure}[!thb]
  \centering
 \includegraphics[width=0.8\columnwidth]{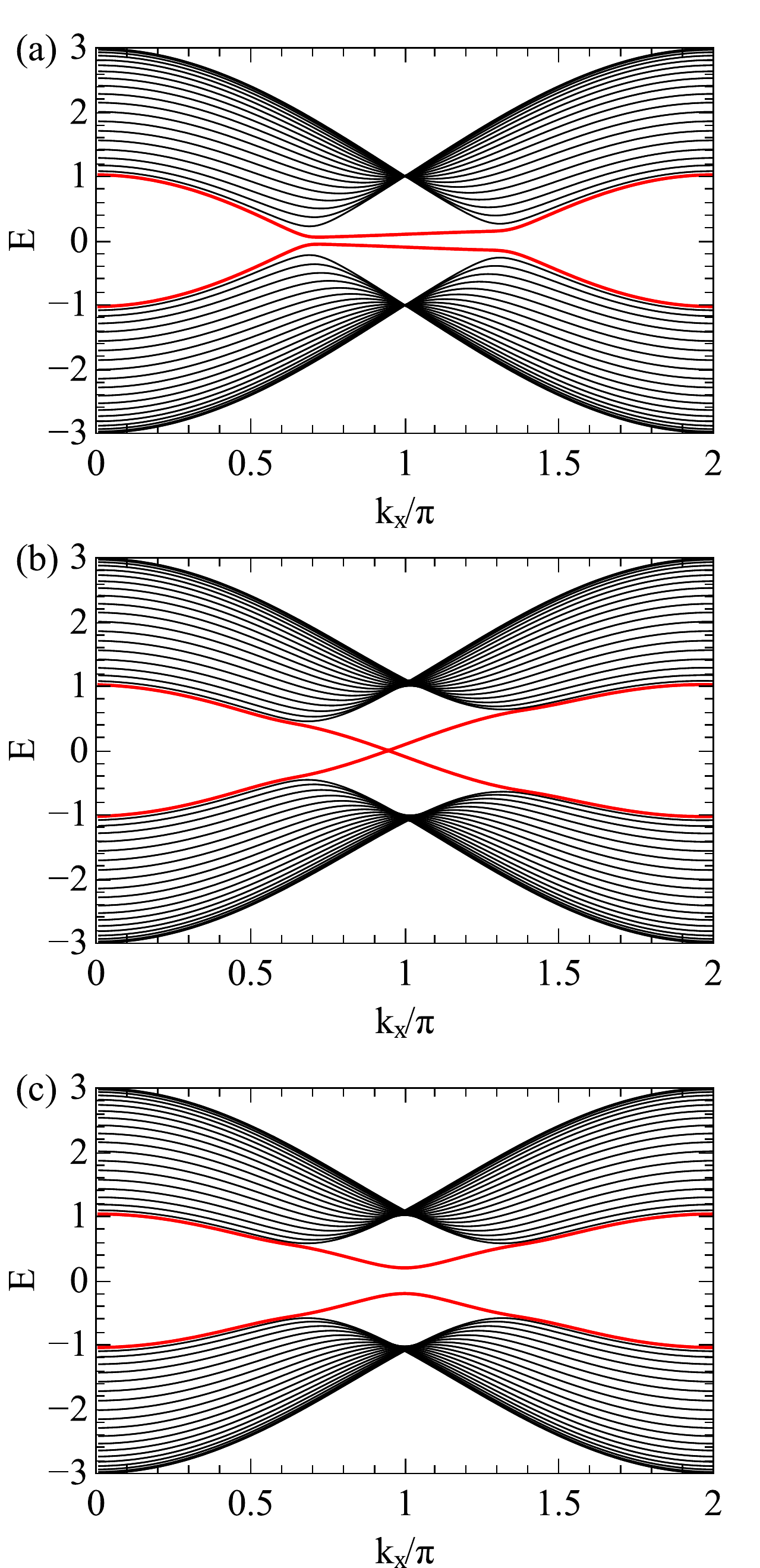}
 \caption{Energy spectra of the trivial AFM$_{z}$ state [in (a)], the topological AFM$_{z}$ insulator [in (b)], and the AFM$_{xy}$ state [in (c)] from finite slab calculations. Black lines denote the bulk states and red lines denote edge states. The topological AFM$_z$ state is characterized by the gapless edge states. See text for detailed information on the parameters.}
\label{fig8}
\end{figure}

\begin{figure}[!thb]
  \centering
  \includegraphics[width=0.8\columnwidth]{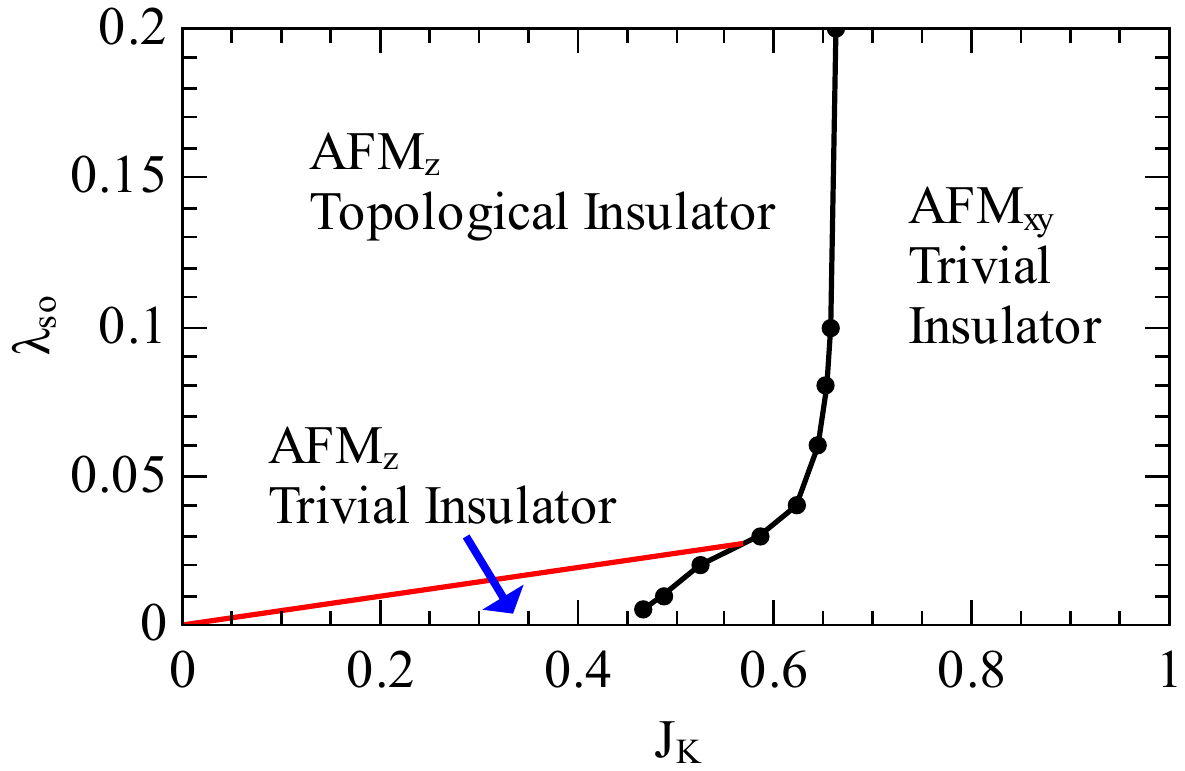}
 \caption{Phase diagram in the $\lambda_{so}$-$J_K$ plane showing the competition of various AFM states. The red line denotes a topological phase transition between the topological trivial and topological nontrivial AFM$_z$ states, and the black curve gives the boundary between the AFM$_z$ and AFM$_{xy}$ states. These two
 states become
equivalent in the limit of $\lambda_{so}\to0$.}
\label{fig9}
\end{figure}

To support the above analysis, we perform calculations of the energy spectrum of the conduction electrons in the AFM$_z$ and AFM$_{xy}$ states, as shown in Eq.(\ref{afmx}) and Eq.(\ref{afmz}), on a finite slab of size $L_{x} \times L_{y}$, with $L_{x}=200$ and $L_{y}=40$. The boundary condition is chosen to be periodic along the $x$ direction and open and zig-zag-type along the $y$ direction. In Fig.~\ref{fig8} we show the plots of the energy spectra with three different set of parameters: (a) $\lambda_{so}=0.01$, $J_{K}=0.4$, $M_{z}=0.5$, (b) $\lambda_{so}=0.0$, $J_{K}=0.4$, $M_{z}=0.5$, and (c) $\lambda_{so}=0.0$, $J_{K}=0.8$, $M_{x}=0.5$, which respectively correspond to the topologically trivial AFM$_{z}$ state, topological AFM$_{z}$ insulator, and AFM$_{xy}$ state. As clearly seen, the gapless edge states only exist for parameter set (b), where the system is in the topological AFM$_z$ state. Note that in this state, the spectrum is asymmetric with respect to the Brilluion zone boundary ($k_x=\pi$), reflecting the explicit breaking of TRS. Based on our analysis and numerical calculations, we construct a phase diagram, shown in Fig.~\ref{fig9}, to illustrate the competition of these AFM states. As expected, the AFM$_z$ state is stabilized for $J_K\lesssim0.7$, and is topological for $J_K<12\sqrt{3} \lambda_{so}$ (above the red line).

\section{Discussion and Conclusion}
We
 have discussed
 the
  properties of various phases in the ground-state phase diagram of the
  spin-orbit-coupled Kondo lattice model on the honeycomb lattice at half filling.
  We
  have shown how
  the competition of SOC, Kondo interaction, and magnetic frustration
  stabilizes
  these phases. For example, in the AFM phase the moments can order either along the $z$-direction or within
  the
  $x$-$y$ plane.
 In our model, the AFM
 order is driven
 by the RKKY interaction, and the competition of SOC and Kondo interaction dictates the direction of the ordered magnetic moments.

Throughout this work, we have discussed the phase diagram of the
model at
half filling.
The phase diagram away from half-filling is also an interesting problem. We expect that the competition between the AFM$_z$ and AFM$_{xy}$ states
persist at generic fillings,
but the topological feature
will not.
Another interesting filling would be the dilute-carrier limit, where a DKSM exists, and can be tuned to a TKI by increasing SOC.\cite{Feng_2016}

In this work we
have
considered a particular type of SOC, which is inherent in the bandstructure of the itinerant electrons. In real materials,
there are also
SOC
terms that involve the magnetic ions.
Such couplings will lead to
models beyond the
current work,
and may further enrich the global phase diagram.

In conclusion, we have investigated the ground-state phase diagram of a
spin-orbit coupled Kondo-lattice model at half-filling.
The combination of SOC, Kondo and RKKY interactions
produces
various quantum phases, including a Kondo insulator, a topological insulator with VBS spin correlations, and two AFM phases.
Depending on the strength of SOC, the magnetic moments in the AFM phase can be either ordered
perpendicular to
or in the $x$-$y$ plane. We further show that the $z$-AFM state is topologically nontrivial for strong and moderate SOC,
and can be tuned to a topologically trivial one via a topological phase transition by
varying
either the SOC or the Kondo coupling.
Our results
shed new light on the
global phase diagram
of
heavy fermion materials.

\section*{Acknowledgements}
We thank W. Ding, P. Goswami, S. E. Grefe, H.-H. Lai,
Y. Liu, S. Paschen, J. H. Pixley, T. Xiang, and G. M. Zhang for useful discussions.
Work at Renmin University was supported by the Ministry of Science and Technology of China, National Program on Key Research Project Grant number 2016YFA0300504, the National Science Foundation of China Grant number 11674392 and the Research Funds of Remnin University of China Grant number 18XNLG24. Work at Rice was in part supported by the NSF Grant DMR-1611392 and the Robert A. Welch Foundation Grant C-1411. Q.S. acknowledges the hospitality and support by a Ulam Scholarship from the Center for Nonlinear Studies at Los Alamos National Laboratory.


\end{document}